\documentclass[runningheads]{llncs}
\usepackage[T1]{fontenc}

\usepackage{tikz}
\usepackage{hyperref}
\usepackage{doi}
\usepackage{booktabs}
\usepackage{rotating}
\usepackage{adjustbox}
\usepackage[utf8]{inputenc}
\usepackage{ltablex}    
\keepXColumns
\usepackage{graphicx}
\usepackage{makecell}
\usepackage{booktabs}
\usepackage{amsmath}

\begin{document}

\title{Exploring Trust Calibration in XAI - The Impact of Exposing Model Limitations to Lay Users}
\titlerunning{Exploring Trust Calibration in XAI - Impact of Exposing Model Limitations}
%
\author{Alfio Ventura\inst{1,2}\orcidID{0000-0003-1639-8001}\thanks{Equal contribution and shared first authorship.}\and
Tim Katzke\inst{1,3}\orcidID{0009-0000-0154-7735}\protect\footnotemark[1] \and
Jan Corazza\inst{1,3}\orcidID{0009-0000-1342-0117} \and
Mustafa Yal\c{c}{\i}ner\inst{1,3}\orcidID{0009-0005-6240-7062}
}
\authorrunning{Ventura et al.}
%
\institute{Research Center Trustworthy Data Science and Security, University Alliance Ruhr, Joseph-von-Fraunhofer-Str. 25, Dortmund, 44227, Germany \and
University of Duisburg-Essen, Bismarckstraße 120, Duisburg, 47057, Germany \and
TU Dortmund University, August-Schmidt-Straße 1, Dortmund, 44227, Germany\\
\email{alfio.ventura@uni-due.de, tim.katzke@tu-dortmund.de}\\
}


%
\maketitle              
\begin{abstract}

Trust calibration---aligning user trust judgment with model capability---is crucial for safe deployment of explainable AI (XAI), yet is often evaluated via global trust ratings detached from objective performance evidence.
We present a preregistered, incentivized between-subject online study (N=418 representative UK sample) on explainable skin-lesion classification that disentangles expectation-setting from experienced performance. Participants completed 15 case evaluations using a fixed XAI panel (malignancy score, reliability score, and saliency map). We systematically manipulated five experimental onboarding conditions varying example-based information and limitation disclosures with five stimulus packages naturally varying observed prediction quality. Calibration was operationalized as the deviation between trust-related judgments (TAIS and case-wise ratings) and objective performance benchmarks for the encountered cases, analysed with hierarchical mixed-effects models.
Only limitation disclosure for case-wise measures reliably impacts trust calibration, and short-term experience did not yield progressive calibration. Further, the experienced package of stimuli explained substantially more variance than the experimental manipulation. However, participants were hard-pressed to differentiate between case-wise perceived trust, trustworthiness, and accuracy estimation.
We discuss implications for designing limitation communication and for measuring and analysing calibration metrics in XAI evaluations.
All study materials and data of this study are publicly available for replication and further academic use.


\end{abstract}

\section{Introduction}

As AI-based decision support services become more widely deployed, model outputs and accompanying explanations are increasingly presented to users who are not domain experts~\cite{DBLP:journals/health/SunAS25,Wongvibulsin2024DermAIApps}. This shifts the human factors problem from supporting expert oversight to supporting appropriate reliance: users must decide when to accept, question, or disregard model advice without the procedural scaffolding that typically constrains interpretation and follow-up actions~\cite{HoffBashir2015,Lee2004,Parasuraman1997}. In high-stakes settings such as healthcare---and particularly in consumer-facing medical applications---both over-reliance (treating the model as authoritative) and under-reliance (discounting useful signals) can be consequential, making trust calibration a safety-relevant requirement rather than a secondary usability outcome~\cite{Lee2004,Parasuraman1997,Skitka2000Accountability,Skitka1999AutomationBias}. 

Much prior work in XAI evaluation therefore examines whether explanations increase trust, perceived usefulness, or user satisfaction~\cite{DBLP:journals/corr/abs-2504-12529,Chromik2021Illusion,Leichtmann2023}. However, higher trust is not inherently desirable: the key requirement is \emph{calibrated} trust, i.e., whether trust-related judgments (and, ideally, reliance decisions) track the system's actual capability under the conditions users experience, i.e., "objective" trustworthiness~\cite{Lee2004,Wischnewski2023,DBLP:conf/fat/ZhangLB20}. Empirically, this distinction is often blurred because trust is frequently measured as a global attitude, while calibration requires linking judgments to objective performance evidence at the level of concrete cases and accounting for how beliefs update over repeated interactions~\cite{Wischnewski2023,RechkemmerYin2022ConfidenceAccuracy,DBLP:conf/fat/ZhangLB20}. Moreover, trust-related constructs can diverge (e.g., perceived trust vs.\ demonstrated reliance), and early framing can anchor later mental models, complicating the interpretation of single-shot trust measurements~\cite{AbbaspourOnari2026Dynamics,Nourani2021AnchoringBias}.

\begin{figure}[t]
    \centering
    \includegraphics[width=\linewidth]{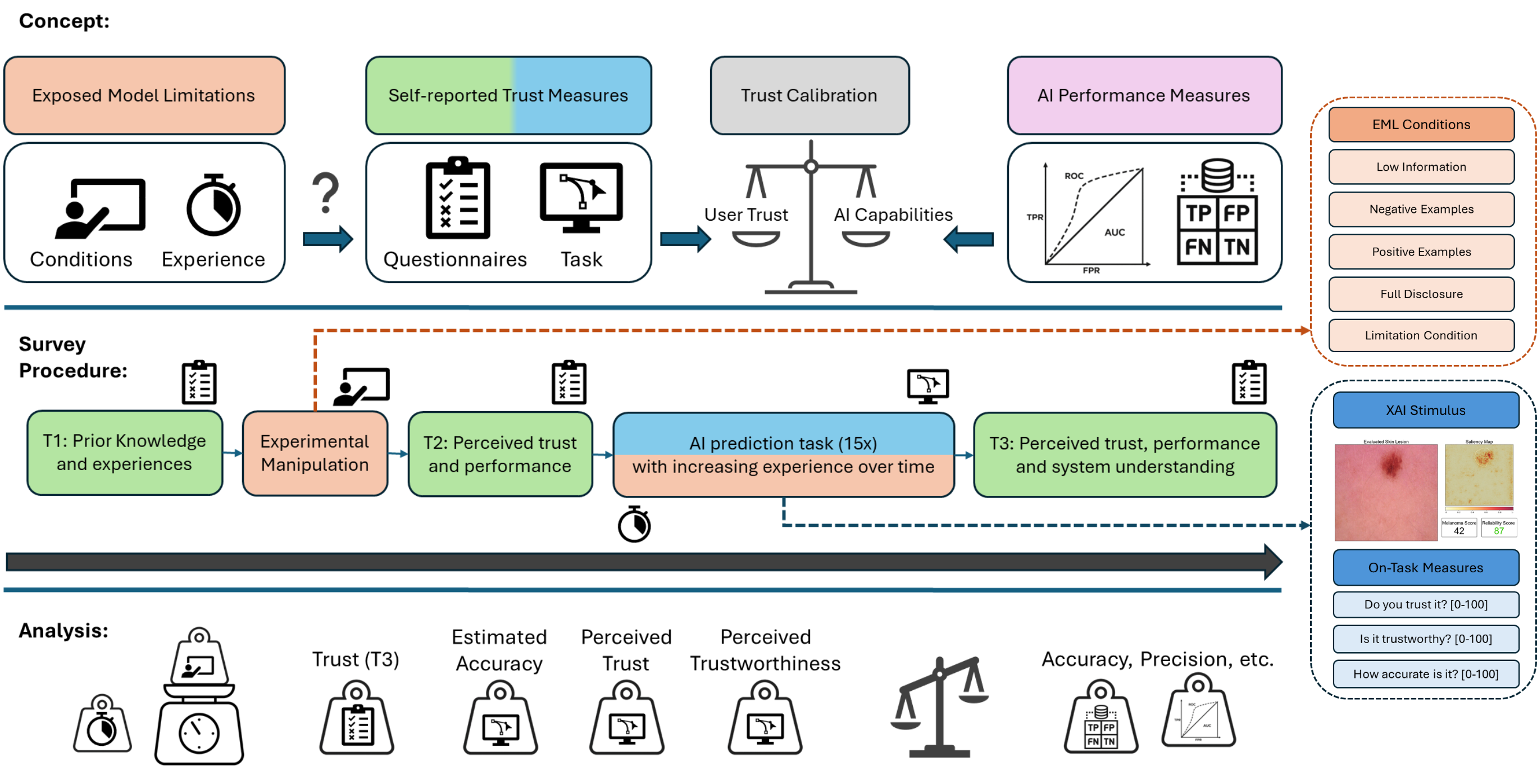}
    \caption{Graphical overview of our overarching study concept, the major survey procedure steps with their key components, and the core analyses performed.}
    \label{fig:graphical_abstract}
\end{figure}

Our research seeks to address these gaps through a preregistered, incentivized user study in the context of skin lesion analysis (\autoref{fig:graphical_abstract}). Participants evaluated a sequence of AI predictions presented with a static XAI interface. We specifically examine how trust calibration is shaped by information and examples shared during onboarding about model performance and limitations, and what users experience about prediction quality during evaluation. To this end, participants were randomly assigned to one of five experimental conditions that varied the disclosure of performance evidence and limitations during onboarding, and to one of five stimulus packages that systematically varied the prediction quality encountered during the task.
Accordingly, this study is guided by the following preregistered research questions and hypotheses:
\begin{description}
  \item[RQ1:] Does the degree of explaining how AI predictions are generated influences calibration of trust-related judgments? Does experience with the system over time improve trust calibration?
  \item[RQ2:] Do users intuitively calibrate their trust-related judgments to one or more "objective" trustworthiness indicators of the AI system?
  \item[RQ3:] Do users intuitively distinguish between different AI evaluation metrics (trust, trustworthiness and accuracy) and to what extent do they differ?
\end{description}

We expect that \textbf{H1:} perceived trust(worthiness) is best calibrated by exposing model limitations, and that \textbf{H2:} perceived trust(worthiness) calibrations should improve over time and experience.
The additional research questions are approached exploratively.

This work contributes to the literature (a) a rich dataset from an interdisciplinary user and AI study, allowing (b) trust calibration to be operationalized as the alignment between trust-related judgments and real-life objective performance benchmarks. We (c) employ sophisticated hierarchical inferential statistics, analysing trust differences, trust calibration, and alignments. Finally (d), we derive various methodological implications for evaluating calibration in future XAI user studies.

\section{Related Work}

\subsection{Trust and trust calibration in Human--AI interaction}
In human factors and human--automation research, the key objective is not trust but enabling \emph{appropriate reliance}: users should consider the helpfulness of automated aids, thereby avoiding misuse (over-reliance) and disuse (under-reliance) (\cite{Lee2004,Parasuraman1997}). This framing motivates a principled distinction between \emph{trust} as a user-side psychological state (willingness to accept vulnerability/delegate) and \emph{trustworthiness} as system-side properties (e.g., competence, reliability)~\cite{HoffBashir2015,Lee2004}. \emph{Trust calibration} then concerns whether trust-related judgments and reliance behavior \emph{track} actual capability under the operating conditions users encounter, rather than merely reporting high or low global trust~\cite{Lee2004,Wischnewski2023}.

Trust perceptions are inherently dynamic and experience-driven: users combine priors (e.g., dispositional propensity, perceived risk) with observed system behavior, updating beliefs after salient successes or failures~\cite{Dietvorst2015,HoffBashir2015,Logg2019}. Such dynamics are safety-relevant as known failure patterns, such as \emph{automation bias}, lead to commission errors (following incorrect recommendations) and omission errors (missing critical events not flagged by the system)~\cite{Skitka2000Accountability,Skitka1999AutomationBias}, with possible fatal consequences. These phenomena imply that effective trust interventions in high-stakes XAI depend on user \emph{recognizing} high- and low-reliability situations and \emph{adapting} reliance accordingly, not on influencing average trust perceptions.

Methodologically, much empirical work still operationalizes trust as a global self-report, whereas calibration is relational (trust vs.\ capability) and time-dependent (learning across interactions). Common instruments include generic trust-in-automation scales~\cite{Jian2000,Koerber2019} as well as newer AI-specific multi-dimensional (e.g., ability, integrity, transparency, vigilance) measures such as the Trust in AI Scale (TAIS) under a global trust factor \cite{TAIS_Wischnewski_Doebler_Krämer_2025}. Complementing attitudes, task-grounded instruments and behavioral measures aim to capture trust-relevant behavior more directly; for instance, the Trust Of Automated Systems Test (TOAST) separates perceived \emph{system understanding} and perceived \emph{system performance} as two factors \cite{Wojton2020}. Recent work within the XAI community further emphasizes that \emph{perceived} trust and \emph{demonstrated} trust can diverge depending on interaction mode and risk framing, strengthening the case for jointly modeling subjective trust and reliance across time~\cite{AbbaspourOnari2026Dynamics}. Consistent with this, a large survey across automation domains highlights substantial heterogeneity in calibration interventions and in how calibration is measured, frequently limiting comparability and interpretation \cite{Wischnewski2023}. Our work follows this methodological critique by explicitly linking repeated trust-related judgments to objective, case-specific performance signals, and by separating onboarding-induced priors from experience-driven updating.

\subsection{Communicating limitations and uncertainty in AI}
A primary lever for calibrated trust is \emph{how capability, uncertainty, and limitations are communicated}---both through onboarding (what users are told) and through interface cues (what users see during decisions). In XAI, explanations can increase perceived transparency and trust without reliably improving decision quality, and may foster misplaced confidence when interpreted as persuasive rationales rather than diagnostic evidence~\cite{Chromik2021Illusion,Leichtmann2023,Wischnewski2023}. Accordingly, many systems augment explanations with explicit performance indicators such as confidence, reliability/uncertainty scores, or stated accuracy~\cite{RechkemmerYin2022ConfidenceAccuracy,DBLP:conf/fat/ZhangLB20}. Evidence from controlled studies suggests that \emph{observed} accuracy during interaction often dominates \emph{stated} accuracy and confidence in shaping both reliance and trust reports, while early cues can anchor subsequent mental models and reliance patterns~\cite{Nourani2021AnchoringBias,RechkemmerYin2022ConfidenceAccuracy}. More recent CHI work argues for instance-level calibration that considers both AI and human correctness likelihood, rather than treating AI confidence as a sufficient proxy for trustworthiness~\cite{Ma2023WhoShouldITrust}.

Beyond passive disclosure, interaction design can reduce over-reliance by introducing scaffolding that preserves user agency. Cognitive forcing functions that require users to first commit to an independent judgment before seeing AI advice reduce over-reliance and can improve hybrid performance even when explanations are available~\cite{Bucinca2021CognitiveForcing}. Complementary design work frames explanation interaction as a training and sensemaking process and proposes engagement- and transparency-oriented design choices to better align reliance with actual capability~\cite{DBLP:journals/computer/NaisehCAJA21,Wischnewski2023}.

At the ecosystem level, documentation proposals such as Model Cards and Datasheets aim to standardize how scope, evaluation evidence, and limitations are disclosed to downstream stakeholders~\cite{Gebru2021Datasheets,Mitchell2019ModelCards}. This issue becomes acute when medical AI outputs reach lay users through consumer-facing tools. For dermatology, a scoping review of mobile applications with AI features reported limited validation and inconsistent transparency about evidence and regulatory status, underscoring risks of miscalibrated user trust in the absence of clear limitation communication~\cite{Wongvibulsin2024DermAIApps}. In parallel, meta-analytic evidence indicates that skin-cancer classifiers can match or exceed non-expert clinicians on curated test sets but exhibit substantial variability across datasets and settings, reinforcing that communication should focus on limitations and uncertainty rather than global "human-level" claims~\cite{salinas_systematic_2024}.

Building on this literature, our study isolates the effects of onboarding disclosures of performance evidence and limitations and experienced prediction quality across repeated cases, while holding a static XAI interface constant; the interface design is informed by SkinSplain, which combines malignancy and reliability cues with a visual explanation signal for skin lesion classification \cite{katzke2025skinsplain}.

\section{User Study Design}
To address the research questions introduced above, we conducted a preregistered, incentivized between-subjects online experiment in which lay users evaluated explainable AI outputs for skin lesion classification \autoref{fig:graphical_abstract}.
\footnote{The preregistration, study materials, data, as well as supplemental materials are publicly available on the
Open Science Framework (OSF): \url{https://osf.io/mkwex}}.
The study crossed two sources of information that plausibly shape calibration: (i) an experimental onboarding manipulation varying how model limitations and performance evidence were communicated (five conditions), and (ii) experienced model performance during the task, operationalized through assignment to one of five stimulus packages.
After standardized scenario briefing and interface familiarization, participants completed a repeated evaluation task consisting of 15 independent cases presented via a static XAI interface. For each case, participants provided case-wise judgments capturing trust-related perceptions (trust in the prediction, perceived prediction trustworthiness, and an estimated prediction quality score), complemented by system-level trust and covariate measures collected before and after the task. In our evaluation, we treat trust calibration as alignment between a participants trust-related judgments and objective performance benchmarks associated with the cases they experienced. A performance-based bonus for accurate quality estimation incentivized careful assessment without directly rewarding higher or lower trust.

\subsection{Stimuli and XAI Cues}

\begin{figure}[t]
    \centering
    \includegraphics[width=0.8\linewidth]{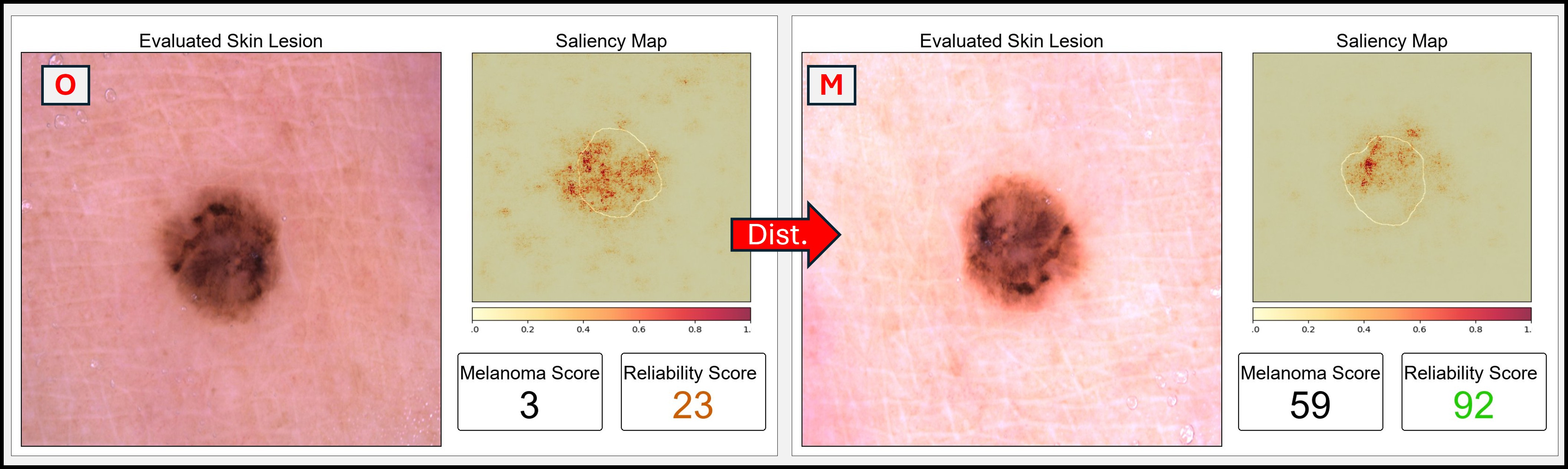}
    \caption{Example prediction presented to participants in the experimental group Limitation Condition introduction. Participants in the Low-Information group did not see such example predictions; the other groups were only presented with the original image on the left without any manipulations.}
    \label{fig:stimuli}
\end{figure}

Each trial in the prediction task presented a single, static stimuli based on XAI cues (Figure~\ref{fig:stimuli}, see also \cite{katzke2025skinsplain}), composed of four elements: (1) a dermoscopic image of a skin lesion, (2) a melanoma score summarizing an AI model's malignancy prediction, (3) a reliability score intended to communicate how dependable the model's prediction is for the given input, and (4) a saliency map visualizing pixel-level importance and human-expert drawn melanoma outlines. 

\paragraph{User-facing XAI cues (per case)}

Our cue set and its visual layout were derived from the SkinSplain framework \cite{katzke2025skinsplain}, but deployed here as fixed stimuli rather than an interactive interface (two examples are given in Figure~\ref{fig:stimuli}). This design preserves the core interpretability and trust calibration cues (numerical prediction, reliability signal, and visual explanation) while keeping the perceptual input constant across participants and trials. 

The melanoma score communicates the prediction output of a binary classifier, intended as an intuitive confidence-like indicator for benign vs. malignant classification. The classifier is based on a fine-tuned EfficientNet-V2-S model~\cite{tan2021efficientnetv2}, trained on a labeled subset of the ISIC Challenge datasets~\cite{ha2020identifying} which focuses only on melanoma and nevus lesions, achieving strong performance on the held-out test data (0.9548 AUC-ROC)~\cite{katzke2025skinsplain}. For our study, this score is scaled to a range of 0 to 100, with higher values indicating greater likelihood of melanoma.

Next, the reliability score operationalizes how aligned the input is with the model's training distribution, i.e., whether the model is "in familiar territory" for this case. It is computed using a layer-wise Deep k-Nearest Neighbors–style procedure that quantifies consistency of latent behavior across layers relative to training neighbors~\cite{DBLP:journals/corr/abs-1803-04765}; lower reliability (scaled to a range of 0 to 100) indicates that the prediction should be interpreted with more caution even when the melanoma score appears decisive.

Saliency maps are generated with an Integrated Gradients approach~\cite{10.5555/3305890.3306024} and gaussian smoothing to highlight regions that most influence the prediction, aiming to support users' qualitative sense-making about what drives the output rather than only how decisive it is. 

\paragraph{Static stimulus generation and packaging}

All stimuli were generated offline from the same underlying technical pipeline described in SkinSplain \cite{katzke2025skinsplain}, using only Skin lesion images from the held-out test set with known ground-truth for prediction labels and skin lesion area segmentation masks for the saliency map attribution. 
We deliberately used a static, survey-compatible rendering of the cue panel instead of interactive exploration, because the study's objective is not to measure exploration behavior, but to test how trust-related judgments track objective performance signals under repeated exposure. 
This stimulus strategy enables a controlled operationalization of "experienced trustworthiness": each participant sees a fixed set of cases with known ground-truth labels (only for us) and corresponding model outputs, making it possible to relate their trust/trustworthiness judgments to the model’s realized performance for the cases they actually encountered. 

For scalability and to support hierarchical modeling of experience, stimuli were organized into five stimulus packages of 15 cases each; participants were assigned to one package of cases presented in randomized order. 
These packages serve as an "experience layer" in the analysis: because package composition determines the mix of correct predictions and error types a participant observes, it induces natural variation in the objective performance signals available for calibration analyses (e.g., package-level accuracy computed against ground truth).
The packages were designed to include one AI prediction each where the AI is very confident and correct, confident and incorrect, and particularly unconfident regardless of correctness.
Furthermore, one AI prediction is selected from each combination in which the information presented to the participants is incongruent. For example, the melanoma score (98) and AI saliency map-expert outline overlap (90\%) agree that this is a confident prediction, but the reliability score is low (11).
Finally, the remaining nine stimuli are selected at random from the hold-out data, with at least one true positive, true negative, false positive and false negative prediction included in each package.

\subsection{Experimental Manipulations}
The experimental manipulations are subtle changes to the AI prediction information and examples presented to users during onboarding.
The aim of the experimental manipulations is to determine whether there is a significant difference between the limitation condition and the other groups receiving progressively more complex information.
All other experimental groups can be considered control groups to a certain extent to determine which additional information impacts perception and behavior.

A total of five experimental groups were implemented: a low-information group presented with no examples (LI), positive examples (PE), negative examples (NE), full disclosure with a positive and negative example (FD), and the limitation condition (LC).
Except for the LI group, all groups received two examples and implications of core prediction information data.
PE information and examples are particularly confident and, therefore, more reliable AI predictions, while NE received particularly unconfident AI predictions.
FD receives one of both confident and unconfident examples.
The LC builds on FD and additionally presents that small distortions (rotation, blurring, brightness adjustments) in the original images may lead to large changes in the AI predictions and XAI cues, even if the underlying skin lesion does not change (see Figure~\ref{fig:stimuli}).
We carried out two manipulation checks to see if a difference could be found between the LC and the other experimental groups.
A Welch two-sample t-test indicates that the LC-manipulation was successful ($p < .001$) in comparison to the other groups ($M_{LC} = 3.99$, $SD_{LC} = 0.60$; $M_{Others} = 3.40$, $SD_{Others} = 0.80$).

\subsection{Evaluation Tasks and Measures}

In total, three questionnaire sets were administered at different points in the online study (see \autoref{fig:graphical_abstract}), in addition to the AI prediction task itself.
An overview of the questionnaires implemented with some core information can be found in (\autoref{tab:Overview_questionniare_table}, for a full documentation please consult the supplemental material.
Respondents were required to answer every item in the questionnaire to avoid missing data.
All items were self-report measures.
\begin{table}[htb!]
\centering
\caption{Overview of the questionnaires and measures employed in the online study.}
\resizebox{\textwidth}{!}{%
  \begin{tabular}{p{6.3cm} c c p{2.3cm}}
\hline
\textbf{Measure} & \makecell{\textbf{Reliability}\\\textbf{Original / Present}} & \makecell{\textbf{M (SD)}\\\textbf{Mdn}} & \textbf{Scale}\\
\hline

\addlinespace[2.0ex]
\multicolumn{4}{l}{\bfseries\fontsize{10.5}{12}\selectfont \textit{T1: Prior to AI Introductions}}\\
\addlinespace[0.2ex]
AI knowledge test (\textit{k}=5) \cite{Leichtmann2023} & $\omega=.50$ / $\omega=.38$ & $Mdn=4$ & 1-of-4\\
ABCDE-Rule for Skin Cancer Detection Knowledge (self-made; \textit{k}=5) & \textemdash / $\omega=.66$ & $Mdn=3.00$ & 1-of-3 \\
Cancer experience (self-made; \textit{k}=1) & \textemdash / \textemdash & \textemdash & 5-point ordinal \\
Propensity to Trust \cite{Koerber2019} (\textit{k}=3) & $\alpha=.75$ / $\alpha=.72$ & $M=3.08 (0.69)$ & 5-point ordinal \\

\addlinespace[2.0ex]
\multicolumn{4}{l}{\bfseries\fontsize{10.5pt}{12pt}\selectfont \textit{T2: After AI Introductions \& XAI Screen}}\\
\addlinespace[0.2ex]
Manipulation Check (\textit{k}=2) & \textemdash / $\omega=.41$ & $M=3.52 (0.80)$ & 5-point ordinal \\
Instructions Check (\textit{k}=2) & \textemdash / \textemdash & \textemdash & forced correct\\
Trust in AI Scale (TAIS; \textit{k}=30) \cite{TAIS_Wischnewski_Doebler_Krämer_2025} & \textemdash / $\omega=.92$ & $M=3.93 (0.47)$ & 5-point ordinal; 6 subscales \\
Perceived System Performance \cite{Wojton2020} (\textit{k}=6) & \textemdash / $\omega=.91$ & $M=5.34 (1.09)$ & 7-point Likert \\
Perceived System Understanding \cite{Wojton2020} (\textit{k}=4) & \textemdash / $\omega=.86$ & $M=5.65 (0.87)$ & 7-point Likert \\

\addlinespace[2.0ex]
\multicolumn{4}{l}{\bfseries\fontsize{10.5}{12}\selectfont \textit{Intermediate: AI Prediction Task (Measured over 15 Independent Stimuli)}}\\
\addlinespace[0.2ex]
Perceived Trust (self-made) & \textemdash / $\omega=.82$ & $M=.70 (.13)$ & 0--100 slider\\
Perceived Trustworthiness (self-made) & \textemdash / $\omega=.79$ & $M=.70 (.13)$ & 0--100 slider\\
Estimated Prediction Accuracy (self-made) & \textemdash / $\omega=.78$ & $M=.72 (.12)$ & 0--100 slider\\

\addlinespace[2.0ex]
\multicolumn{4}{l}{\bfseries\fontsize{10.5}{12}\selectfont \textit{T3: After Prediction Task}}\\
\addlinespace[0.2ex]
Trust in AI Scale (TAIS; \textit{k}=30) \cite{TAIS_Wischnewski_Doebler_Krämer_2025} & \textemdash / $\omega=.93$ & $M=3.89 (0.50)$ & 5-point ordinal \\
Perceived System Performance \cite{Wojton2020} (\textit{k}=5) & \textemdash / $\omega=.84$ & $M=5.42 (0.83)$ & 7-point Likert \\
Perceived System Understanding \cite{Wojton2020} (\textit{k}=4) & \textemdash / $\omega=.87$ & $M=5.62 (0.97)$ & 7-point Likert \\
Familiarity with AI for Skin Cancer Detection (self-made) & \textemdash / $\omega=.88$ & $M=1.53 (0.91)$ & 5-point ordinal \\
Familiarity with ''SkinVision'' App (self-made) & \textemdash / \textemdash & $Mdn="No"$ & Yes/No + open text question \\
Gender perception (self-made; \textit{k}=1) & \textemdash / \textemdash & $M=4.03 (2.88)$ & 7-point ordinal\\
Demographics (sex, age, education) & \textemdash / \textemdash & \textemdash & \textemdash \\
\hline
\end{tabular}
}
\begin{flushleft}\footnotesize
\textit{Note.} A dash (\textemdash) indicates not applicable 
or not reported; \textit{k} represents the number of items in the measure; \(\alpha\) = Cronbach's alpha; \(\omega\) = McDonald's omega. 
\\
\label{tab:Overview_questionniare_table}
\end{flushleft}
\end{table}
Before the participants were informed in detail about the importance of AI in the medical sector and the present AI system, we collected some control variables and basic data.
We tested prior knowledge of AI with a validated AI knowledge test (\cite{Leichtmann2023}), prior skin cancer detection knowledge in terms of the ABCDE rule (\cite{NationalCancerInstitute_ABCDE}), as well as personal experience with cancer.
Finally, we assessed the general propensity to trust towards AI systems as generalized baseline trust.

After participants received detailed information about the AI scenario, they gave an initial system-trust assessment. To assess perceived trust in the AI system, we employed the TAIS (\cite{TAIS_Wischnewski_Doebler_Krämer_2025}) for its psychometric strength and comprehensiveness.
We also collected perceived system performance as control variables and covariates for trust perceptions (\cite{Wojton2020}).

The AI prediction task measures consisted of evaluating perceived trust, perceived trustworthiness, and AI system prediction accuracy estimates.
After task completion, we collected the same data as in T2, in addition further control variables familiarity with AI systems for skin cancer detection in general, as well as familiarity with “SkinVision” specifically.

For the later analyses, these self-reported questionnaire and on-task measurements were compared with established AI performance measurements as "objective" trustworthiness indicators. Specifically, the system-wide accuracy, communicated to the participants in intuitive terms during onboarding, as well as the experienced accuracy, precision, recall, and F1 score, depending on the  AI prediction performance for that specific stimulus package.

\subsection{Study Execution}

\paragraph{Assembling the Online Study}
Data collection took place from 10th October 2025 to 14th October 2025.
All participants who (1) were 18+ years old, (2) were fluent in English, (3) were located in the UK, and (4) had no colour vision deficiencies were eligible to participate on (5) a desktop PC with (6) audio and video capabilities.
All participants consented to participate in the online study and were debriefed afterwards.
All participants created an individualised code word to request future data deletion.

The online study consists of various components (see \autoref{fig:graphical_abstract}).
Participants were randomly assigned to one of the five experimental groups for the information intervention and one of the stimulus packages for the AI prediction task.
The online study employed anti-bot (reCAPTCHA v.2) and anti-generative AI measures (AI trap question at the end), as well as eligibility, two manipulation, two comprehension, and three attention checks (successfully passed by 97.6\%).

Average participation time was 33.55 minutes (10 - 73, $SD$ = 11.67; no difference between stimuli packages [$p$ .974], minor inconsistent differences between experimental manipulations [$p$ .011]).
Participants received £7.20 (£12 per hour) compensation, based on the British minimum wage at the time of the survey.
The top 10\% of participants were able to earn a bonus payment of £1.50, estimating the AI system's prediction accuracy across the 15 experienced AI predictions.

\paragraph{Sample description}
We calculated multiple power analyses that triangulated to a sample size of $n$ = 400 participants (details can be found in the preregistration).
We collected a final sample of 418 eligible submissions by participants based in the UK, employing the online panel provider Prolific.
The sample is stratified by sex (212 identify as female, 202 as male, and 2 as non-binary), age (18 - 78, $M$ = 46, $SD$ = 23.8), and region of residence in the UK.
As with many online panel samples, the present sample is rather well-educated ($Mdn = Bachelor's degree$, 66\% completed a bachelor's degree or higher academic degree).

\paragraph{Information Provided to Participants}
Participants were given extensive introductory information on why AI in medical applications is an important field of innovation.
First, data and statements from the EU were used to communicate the relevance (\cite{EURelevance2021,HorizonGrandViewResearch2023}) as well as opportunities (\cite{EURelevance2021,HorizonGrandViewResearch2023,salinas_systematic_2024}) juxtaposed with risks.

To support participants’ assessment of the AI predictions, we introduced and summarized the ABCDE rule (Asymmetry, Border, Colour, Diameter, Evolving) for skin cancer detection (section 0:36–1:28 of \cite{NationalCancerInstitute_2014}; \cite{NationalCancerInstitute_ABCDE}).
We then explained the visual AI system and the information provided by the prediction interface.
Next, we presented the experimentally manipulated information and examples.
Finally, we added contextual information common to all groups: AI and humans may use very different methods to arrive at their predictions, but this does not affect the validity of the predictions.

After providing this extensive information, we reminded the participants of the task and reiterated the incentive.
To proceed with the study, two questions on the participants' main task and the possible bonus payment had to be answered correctly.
At the end of the instructions, participants had the opportunity to familiarize themselves with the GUI, task, and measures with an example, being able to rotate through different AI predictions (a $\textit{Mdn} = 13$ number of rotations were performed).

\section{Results}
To test the hypotheses on trust calibration and accuracy estimation, we constructed independent hierarchical linear mixed models with fixed and random effects (random intercept, fixed slope).
These models follow a nested structure in which the participants' results are first nested into the experienced stimulus packages, which in turn are nested into the experimental conditions to account for individual trust variance that depends on the corresponding group affiliations.
The models establish the extent to which the experimental manipulations lead to differences in the dependent variables a) final AI system trust assessment (TAIS at T3), b) task trust assessment (combined and aggregated on-task trust and trustworthiness assessment), and c) task accuracy estimation based on the raw data, as well as d) in the calibration of these three measures.
All models were constructed and tested step by step, variable by variable, but only the most central models are reported.
Since the experimental manipulation is categorical, it must be introduced into the linear models as a dummy-coded variable, which necessitates statistical comparisons between the chosen experimental manipulation relative to the others.
The comparison group is specified for all results discussed and will be either the LI control group (answering: is there an effect of the manipulations?) or the LC (answering: is this manipulation the strongest?).

\subsection{Describing Core Dependent Variables}
\begin{figure}[tb]
    \centering
    \includegraphics[width=0.9\linewidth]{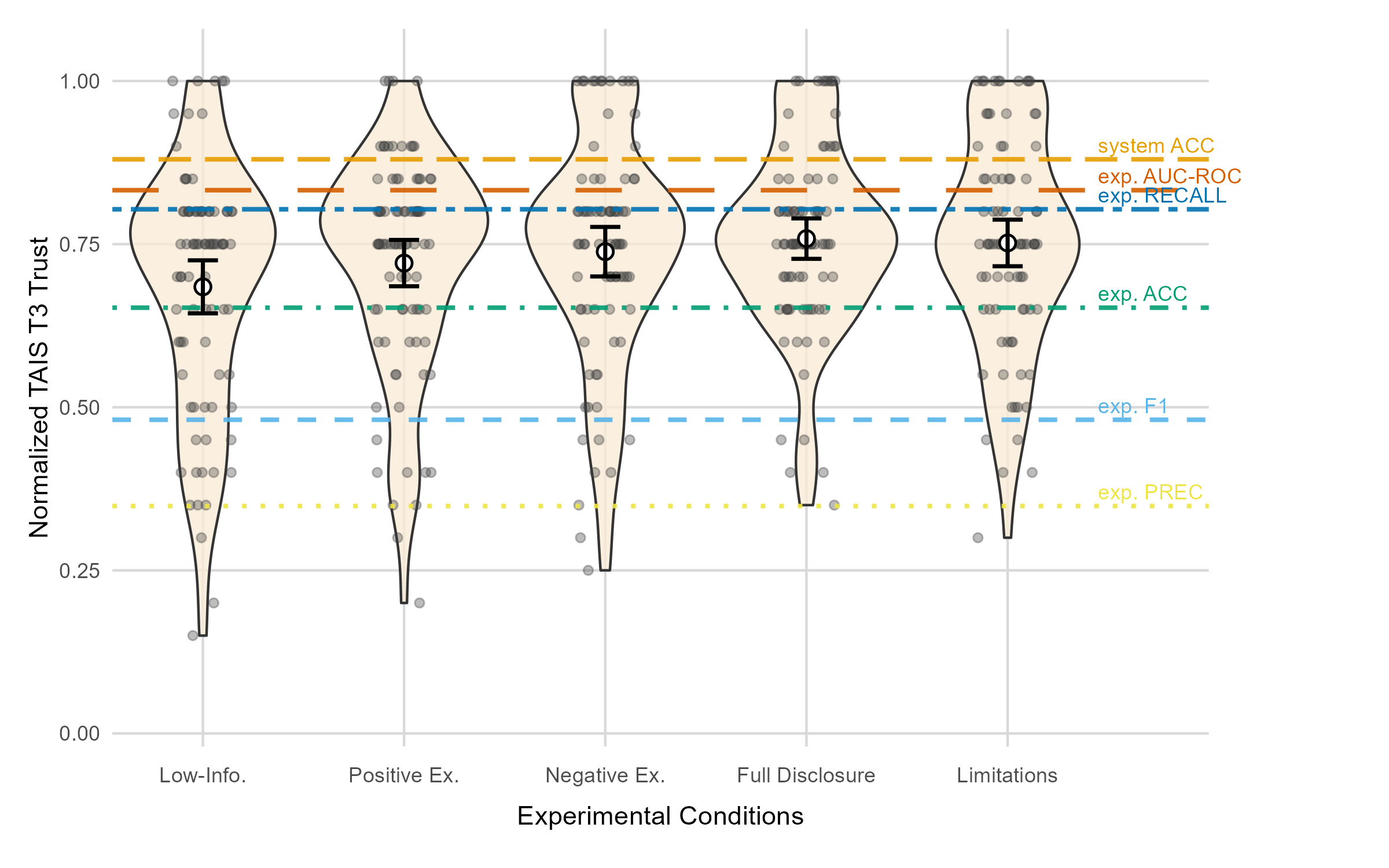}
    \caption{TAIS T3 trust assessment by experimental condition. The "objective" trustworthiness indicators are added as reference lines, either as general system-wide indicators or as mean experienced indicators based on our stimulus packages.}
    \label{fig:raw_TotalTrust}
\end{figure}

\autoref{fig:raw_TotalTrust} presents the raw TAIS T3 data normalized to a 0-1 scale. Trust towards the AI system averages 0.73 and can therefore be considered medium-high.

For each of the fifteen stimuli assessed by each participant, they were asked to rate their trust in the prediction, the perceived trustworthiness of the prediction, and an estimation of the accuracy of the prediction.
The correlation between the three variables is very high at $r_{T-TW} = .93$, ($p_{T-TW} < .001$; $r_{T-EA} = .87$, ($p_{T-EA} < .001$; $r_{TW-EA} = .91$, ($p_{TW-EA} < .001$).
Due to the near-complete overlap between the two trust(worthiness) measures, we computed a combined indicator that we apply to all on-task analyses.
\autoref{fig:raw_taskTotalTrust_and_estimatedAccuracy} presents the raw combined on-task trust as well as accuracy estimation (AE) data.
Trust towards the AI system averages 0.70, AE 0.72. 
Both can therefore be considered medium-high.

\begin{figure}[tb]
    \centering
    \includegraphics[width=0.9\linewidth]{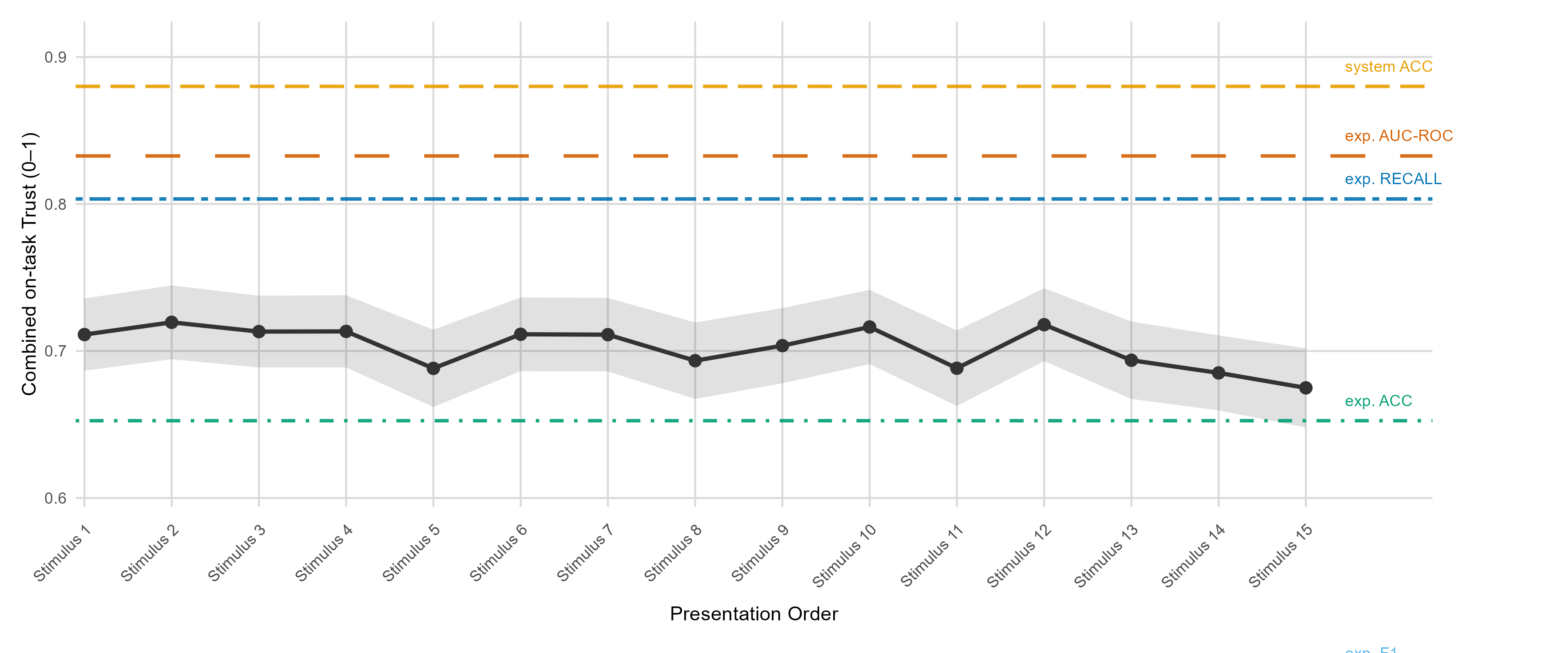}
    \caption{Combined trust and trustworthiness on-task assessment over time. The "objective" trustworthiness indicators are added as reference lines, either as general system-wide indicators or as mean experienced indicators based on our stimulus packages. The shaded band shows the ± 95\% CI around the measures.}
    \label{fig:raw_onTask_longitudinal}
\end{figure}

Finally, \autoref{fig:raw_onTask_longitudinal} presents the raw combined on-task trust data longitudinally over the course of the prediction task.

\subsection{Modeling and Analysing Changes in Dependent Variables}

For each linear mixed-effects model, we report two types of explained variance: marginal variance reflects variance explained by fixed effects, whereas conditional variance reflects total variance explained by both fixed and random effects.
Below, we explain the core analytic elements in detail whenever they first appear.

For the first model comparisons (\autoref{tab:generaltrust_overall_combined}, \autoref{tab:ontasktrust_overall_combined}, \autoref{tab:estimatedAccuracy_overall_combined}), introducing stimulus packages as a random effect (Model 1's) explains low 0.2\% of total variance for the TAIS, medium-large 6.4\% for on-task trust, and 6.8\% AE.

Adding the experimental manipulation as a random and fixed effect to the base models (Model 2's) additionally explains 2.4\% of the total variance ($R^2_c$ = 2.6\%) for TAIS, 5.4\% ($R^2_c$ = 11.8\%) for on-task trust, and 5.0\% ($R^2_c$ = 11.8\%) for AE.
Introducing the stimulus package and experimental conditions lead to statistically significant improvements in the model fit and total variance explained for on-task trust and accuracy estimation, but not for TAIS.
This means that the stimulus package, as well as the experimental condition, explains linear changes at the cluster level (marginal variance; i.e. some different trust values in experimental conditions between stimulus packages) as well as in total (i.e. some differences in experimental conditions independent of stimulus packages) for on-task trust and AE, but not for TAIS.
Furthermore, additional total variance is explained by the random effects on top of the fixed effects (total variance; i.e. some differences between stimulus packages) for on-task trust and AE.

Analysing the fixed effects for TAIS, compared to LI (Model 2a), both LI vs. LC and LI vs. FD predict statistically significant trust.
Compared to LC (Model 2b), there is only a statistically significant difference with LC vs. LI.
For on-task trust, LI vs. LC differ statistically significantly from each other, for AE, it is LI vs. LC as well as LC vs. PE.

Moving on from models 1 and 2, we add some core control variables to further models to isolate the influence of the experimental condition on trust by holding other unsystematic differences between the groups constant.
We decided to add the control variables AI knowledge, skin cancer knowledge, and propensity to trust because they were collected before our AI system was introduced to the participants.
Further control variables added are the individual effort, operationalised by study completion time, and familiarity with similar AI-for-Skin-Cancer-detection systems.
Of these control variables (Model 3's), for TAIS only propensity to trust statistically significantly predicts later trust, for on-task trust and AE it is propensity to trust as well as familiarity.
The model fit and conditional variance explained increase significantly for TAIS ($\Delta R^2_c$ = 14.5\%), and on-task trust ($\Delta R^2_c$ = 16.5\%), but not for AE ($\Delta R^2_c$ = -2.1\%).

When the experimental manipulation is added to these models (Model 4's), it only explains additional statistically significant total variance for AE ($\Delta R^2_c$= 4.8) and on-task trust ($\Delta R^2_c$= 4.9), but not for TAIS ($\Delta R^2_c$= 1.4).
The statistically significant differences observed between the experimental manipulations become insignificant for TAIS after including the control variables, indicating that the differences arise from a priori group differences.
For AE, only LC vs. PE as well as propensity to trust becomes insignificant, but LI vs. LC stays.

\subsection{Modeling and Analysing Dependent Variables as a Measure of (Trust) Calibration}

\begin{figure}[t!]
    \centering
    \includegraphics[width=0.9\linewidth]{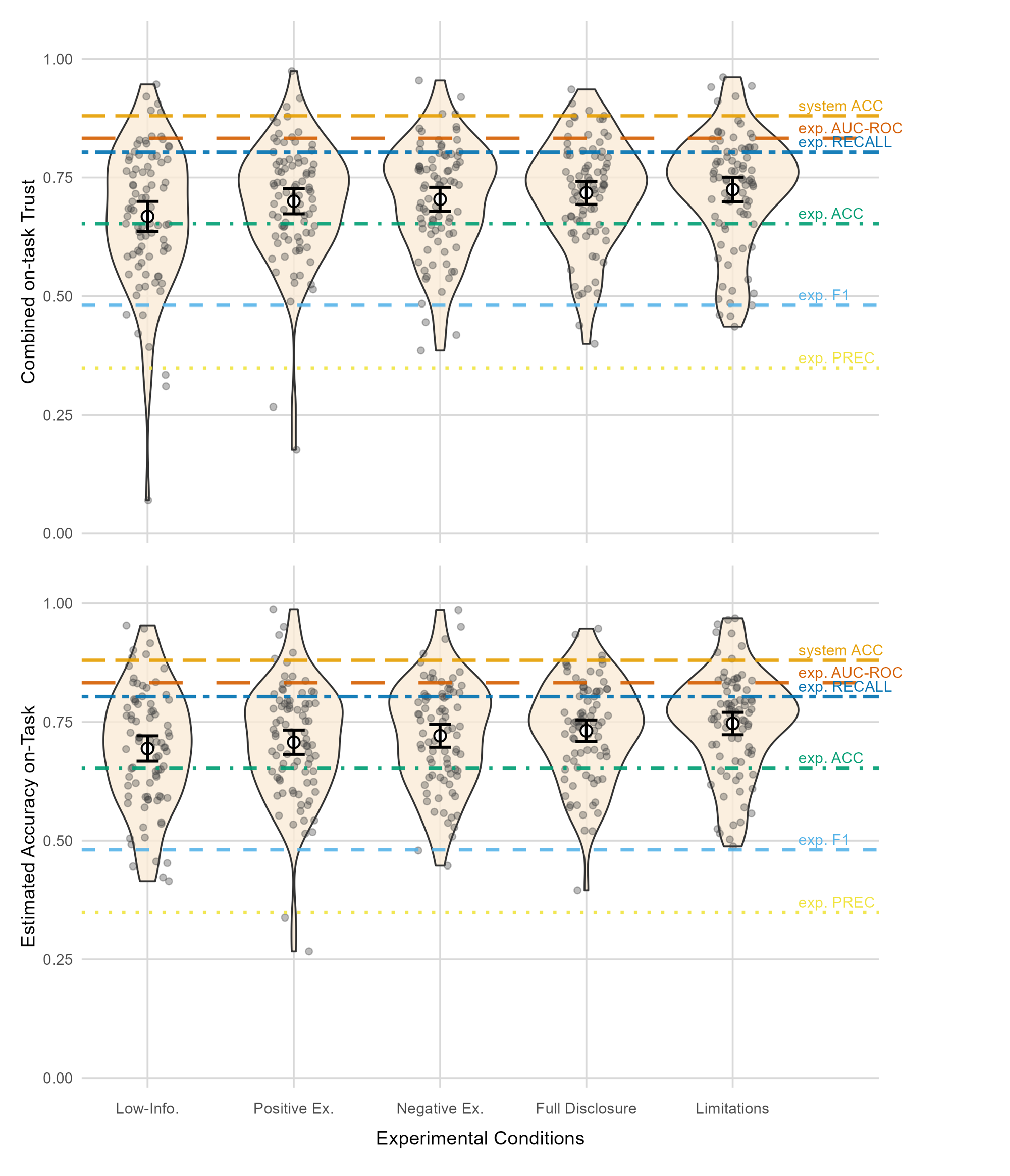}
    \caption{On-task assessments of combined trust and trustworthiness (top) and estimated accuracy (bottom) by experimental condition. The "objective" trustworthiness indicators are shown as reference lines, either as system-wide indicators or as mean experienced indicators based on the stimulus packages.}
    \label{fig:raw_taskTotalTrust_and_estimatedAccuracy}
\end{figure}

Here, we treat TAIS, on-task trust, and AE as distances from the AI system’s “objective” trustworthiness indicators, yielding tables analogous to those in the appendix.
Statistical details are provided in the supplemental materials, as the analyses produce 3×(6+1) tables.
We focus here on the overarching results.

Overall, introducing the stimulus package as random effect (Model 1's) explains much variance for TAIS ($Mean R^2_c$ = 12.4\%), and substantial variance for on-task trust ($Mean R^2_c$ = 30.6\%) as well as AE ($Mean R^2_c$ = 33.1\%) as a measure of calibration.
For TAIS, substantially above average explained variance is observed for PREC, ACC, RECALL and f1 scores.
For on-task trust, it is for PREC, RECALL and f1 scores, for AE only RECALL and f1 scores.

Adding the experimental manipulation as random and fixed effects (Model 2's) additionally explains on average 1.5\% of the total variance for TAIS, 3.2\% for on-task trust, and 3.1\% for AE.
For TAIS, introducing the manipulations does not lead to statistically significant improvements in the model fit and, therefore, in the total variance explained.
For on-task trust 3/6, and AE 4/6 cases, introducing the experimental conditions led to statistically significant improvements.
Furthermore, for TAIS and on-task trust in 3/6, for AE in 4/6 analyses, a statistically significant difference between LI and LC can be observed.
In addition, for AE in 4/6 analyses PE vs. LC, in 1/6 NE vs. LI, research statistical significance.

In comparison to the analyses in the previous section, the stimulus packages explain far more variance in trust-calibration than in raw values. 
The average variance explained by the experimental conditions remains comparable for TAIS, and decreases for both on-task measures.
Therefore, participants seem to pick up on differences in stimulus package prediction quality, which affects their assessment.
This is especially true for the on-task measures.

In the next model comparisons (Model 3's), core control variables are again added to isolate the influence of the experimental condition on trust.
The control variables add substantial variance explained for TAIS ($Mean R^2_c$ = 21.8\%), with only propensity to trust consistently reaching statistical significance as a predictor (5/6 analyses).
They add even more variance for on-task trust ($Mean R^2_c$ = 33.6\%) and AE ($Mean R^2_c$ = 35,2\%) with propensity to trust (4/6) and familiarity with visual AI (3/6) consistently reaching statistical significance for on-task trust, and only familiarity with visual AI (3/6) for AE.
After adding all variables to the models (Model 4's), for both on-task trust (2/6) and AE (3/6) adding the experimental manipulations leads to one less statistically significant difference.
For TAIS, the LI vs. LC become non-significant after adding the control variables to the models.
For on-task trust (3/6) and AE (4/6) LI vs. LC stay statistically significant, but the only other differences that stay significant for AE are PE vs. LC (4/6).

\subsection{Comparing Dependent Variables to "Objective" Trustworthiness Indicators}
In these analyses, the final model 4's (control variables, experimental manipulations, and stimulus packages) analyses from \autoref{tab:generaltrust_overall_combined}, \autoref{tab:ontasktrust_overall_combined} and \autoref{tab:estimatedAccuracy_overall_combined} are recycled.
We compute estimated marginal means (EMMs; emmeans) for each experimental condition from the previously reported mixed-effects model 4's and compare each EMM with the six "objective" trustworthiness indicators using Wald t-tests.
P-values were Holm-adjusted within the analyses for each dependent variable to prevent alpha-error accumulation.
Results are reported as differences between the EMMs and the indicators collapsed between stimulus packages.
Detailed inferential statistics comparing the "objective" indicators to the dependent variables are documented in the supplemental material.
For visual intuition of the results, all included figures present the raw data (not EMMs) as comparison with the indicators.

The analyses relative to the TAIS show that there is an inclusion of the indicator in the 95\% confidence interval after adjusting p-values (i.e. no statistically significant differences) with RECALL (4/5 experimental conditions; $R^2_c$ = 24.0\%) and experienced ACC (only LI; $R^2_c$ = 18.3\%).
For the on-task trust measure, there is a statistically significant overlap only with experienced ACC (all experimental conditions; $R^2_c$ = 24.1\%).
Same applies to AE and ACC (4/5 experimental conditions; $R^2_c$ = 30.0\%)
The mean total variance explanations of the statistically significantly overlapping models are average or below average, raising doubts about the validity of an actual non-random overlap of the variables and indicators.
Accordingly, the present results indicate that the greatest correspondence between the dependent variables and the “objective” trustworthiness indicators exists for experienced ACC and RECALL.

\subsection{On-Task Trust Evolving over Experience}

Subsequently, we consider how on-task trust develops through experience with the 15 experienced stimuli.
For this analysis, additional random-effect variables are required for the linear mixed effects models: the stimulus order and the interaction between stimulus order and the individual (\autoref{tab:ontasktrust_overTime}).
In model 1, the stimulus order is the only random effect.
In this model, the stimulus order has a statistically significant negative effect on on-task trust.
On-task trust, therefore, decreases over time and experience.

In model 2, the experimental manipulation is introduced as a mediating and control variable.
The experimental manipulation does not, on its own, explain changes in trust over time, but it does lead to the stimulus order no longer explaining any statistically significant variance.
This indicates that changes in on-task trust are no longer explained by experience over time.

We also attempted to fit trust calibration models relative to the ‘objective trustworthiness’ indicators for this longitudinal on-task data.
Unfortunately, we were unable to obtain reliably compiled model-fit estimates, a clear indication of overfitting.
Consequently, we refrain from reporting such results.

\subsection{Modeling Changes in Variance by Experimental Manipulations}
When examining the raw on-task data (\autoref{fig:raw_taskTotalTrust_and_estimatedAccuracy}), a difference in variance between the experimental conditions can be visually detected.
To formally test this visual observation statistically, we performed classic mean-centred Levene tests for the homogeneity of variances.
No statistically significant differences in group variances were found for either the raw values of the TAIS ($F(4, 413) = 2.01$, $p = .09$), on-task trust ($F(4, 413) = 1.47$, $p = .210$) or accuracy estimation ($F(4, 413) = 0.87$, $p = .481$).
The same applies to EMMs of the linear mixed-effect model 4's of the TAIS ($F(4, 413) = 1.18$, $p = .319$), on-task trust ($F(4, 413) = 1.30$, $p = .270$) or accuracy estimation ($F(4, 413) = 1.11$, $p = .352$).
These statistically insignificant tests indicate that the experimental conditions did not affect the variances of the measures.

\section{Discussion and Outlook} 
We conducted a user study in which participants interacted with a visual AI skin-cancer prediction system based on SkinSplain \cite{katzke2025skinsplain} in a repeated, incentivized prediction task.
We experimentally manipulated onboarding and limitation disclosures and examined effects on trust-related judgments using validated questionnaires and case-wise assessments.
We analysed the data using linear mixed-effects models (mean and variance effects) and a calibration metric relative to “objective” trustworthiness indicators.

\paragraph{Interpretation of Main Findings}
Below, we summarize the main findings by research question.
Related to \textbf{RQ1}, \textbf{H1} is partly confirmed.
We found no significant differences in trust or calibration between the example conditions (positive, negative, both/full disclosure) and either the low-information (LI) or limitations (LC) group.
In contrast, LI and LC showed statistically significant, but inconsistent, mean differences on both on-task measures, but not on TAIS, suggesting TAIS is less sensitive to onboarding manipulations than case-wise measures.
Interestingly, PE and NE did not differ, suggesting that example valence did not affect trust-related judgments or calibration at all.
For \textbf{RQ1} \textbf{H2}, we found no evidence that trust calibration developed with task experience.
Notably, the assigned stimulus packages accounted for more variance than the experimental manipulations, suggesting that the specific prediction cases participants encountered were more influential than the experimental onboarding manipulation and its content.
Overall, example-based information and its valence had little impact on trust-related judgments or calibration, whereas limitation disclosure may matter depending on operationalization and methodology.
Most influential, however, was participants’ hands-on experience with the application.

Regarding \textbf{RQ2}, the trust-related judgments seem to align most consistently with ACC and RECALL.
However, our statistical models for ACC and RECALL do not consistently explain the greatest variance among the indicators.
This suggests that the overlap may reflect chance-level alignment rather than participants intuitively picking up on corresponding XAI cues.

Finally, \textbf{RQ3}, differences between trust-related judgments and operationalisations can be observed.
Our analysis models explain more variance for the case-wise measures than for TAIS, suggesting they better match the present study design.
Concurrently, it is unclear how differentiated our on-task measurements actually are perceived to be, because all three measures correlated substantially.
However, the three on-task measures correlated strongly, indicating limited discriminant validity and that participants (even a representative British sample with strong language skills) may not reliably distinguish the underlying concepts without additional guidance.
This, in turn, raises the question of what may predict our core variables.
Evidently, participants’ trust-related judgments were not simple reports of the globally communicated accuracy from the information materials, as the observed pattern deviated from it.
Furthermore, the participants' trust-related judgments are definitely and distinctly influenced by the AI system presented, as the propensity to trust and familiarity with related AI systems remain the most consistently statistically significant predictors in our analyses, yet the variance explained rarely exceeds one-third.

\paragraph{Implications}
According to our findings, examples and information provided in advance do not ensure that users will develop adequate trust-related judgments in existing AI applications.
However, even experience with the AI application appears not to always contribute to trust calibration. At least not in a small time span under experimental conditions.
We conclude that calibrating users’ trust and reliance, alongside robust assessments of trustworthiness, is challenging and requires substantial user involvement.
However, more important than the information participants receive seems to be the individual work samples users experience during actual application use.
Furthermore, it remains unclear whether it is possible to identify shortcuts or approximations that enable professionals to assess in advance whether users would adequately assess the trustworthiness of an AI system and in which direction their trust is likely to develop.

\paragraph{Limitations}
Like any study, ours has limitations. The experiment used an artificial scenario that may not fully generalize to real-world settings.
Although we approximated a first-time AI interaction and incentivized participation, we cannot fully capture the intrinsic motivation that typically drives deeper engagement.
Moreover, while skin-cancer detection is an important use case, it may feel remote for many participants and might not be voluntarily delegated to AI outside the study context.
At the same time, remote, automated preventive check-ups may become more common, making familiarity with AI-based preventive care increasingly relevant.

\paragraph{Future work directions}
First, our study is an initial step toward testing whether explicitly communicating AI limitations supports trust calibration and helps users approximate AI trustworthiness.
Second, we only begin to examine whether users rely on internal cues that align with "objective" trustworthiness indicators. If such alignment exists, it could inform which indicators are most appropriate to represent AI prediction quality. Future work could experimentally manipulate these indicators in the stimulus material to enable stronger causal conclusions.
Third, more user-centered research is needed to clarify how people interpret common “trust calibration” constructs, especially given conceptual disagreement in the literature and potentially limited intuitive distinction between “trust” and “trustworthiness” \cite{Wischnewski2023}. Finally, our dataset supports additional analyses (e.g., effects of prediction incongruence on trust-related judgments, TAIS subscales, and perceived system understanding/performance on effort and calibration), and we invite interested scholars to explore further interesting questions based on it.

\begin{credits}
\subsubsection{\ackname}
This research is funded by the Research Center Trustworthy Data Science and Security (\url{https://rc-trust.ai}), one of the Research Alliance centres within the University Alliance Ruhr (\url{https://uaruhr.de}), and approved by the Ethics Committee of Computer Science at the University of Duisburg-Essen.

\subsubsection{\discintname}
The authors have no competing interests to declare that are relevant to the content of this article. 

\subsubsection{Declaration on Generative AI}
During the preparation of this article, we used the ChatGPT 5.2 model from OpenAI to find additional related work and for language edits, aiming to enhance readability.
After using this service, the authors reviewed and edited the content as needed and take full responsibility for the manuscript's content.

\end{credits}

\begingroup
\bibliographystyle{splncs04}
\bibliography{xai_refs}
\endgroup

\begin{sidewaystable}[p]
    \centering
    \caption{Linear Mixed Effect Models predicting TAIS T3 Trust Measures}
    \label{tab:generaltrust_overall_combined}

    \adjustbox{max width=\textheight, max totalheight=\linewidth, keepaspectratio}{
        \begin{tabular}{lcccccccccccc}
\toprule
\textbf{Model component} & \multicolumn{2}{c}{\textbf{TAIS-M1: Only Packages}} & \multicolumn{2}{c}{\textbf{TAIS-M2a: LI vs Others}} & \multicolumn{2}{c}{\textbf{TAIS-M2b: LC vs Others}} & \multicolumn{2}{c}{\textbf{TAIS-M3: Packages + Controls}} & \multicolumn{2}{c}{\textbf{TAIS-M4a: LI vs Others}} & \multicolumn{2}{c}{\textbf{TAIS-M4b: LC vs Others}} \\
\cmidrule(lr){2-3}\cmidrule(lr){4-5}\cmidrule(lr){6-7}\cmidrule(lr){8-9}\cmidrule(lr){10-11}\cmidrule(lr){12-13}
\midrule
\textbf{Random Effects} & \multicolumn{2}{c}{\textbf{SD}} & \multicolumn{4}{c}{\textbf{SD}} & \multicolumn{2}{c}{\textbf{SD}} & \multicolumn{4}{c}{\textbf{SD}} \\
\addlinespace[0.25em]
Stimulus Package (intercept) & \multicolumn{2}{c}{0.008} & \multicolumn{4}{c}{0.007} & \multicolumn{2}{c}{0.012} & \multicolumn{4}{c}{0.009} \\
Experimental Manipulation (intercept) & \multicolumn{2}{c}{} & \multicolumn{4}{c}{<.001} & \multicolumn{2}{c}{} & \multicolumn{4}{c}{0.015} \\
Residual & \multicolumn{2}{c}{0.170} & \multicolumn{4}{c}{0.168} & \multicolumn{2}{c}{0.156} & \multicolumn{4}{c}{0.155} \\
\addlinespace[0.6em]
\midrule
\textbf{Fixed Effects} & \shortstack[c]{\textbf{$b$}\\\textbf{(SE)}} & \shortstack[c]{\textbf{$t(df)$}\\\textbf{$p$}} & \shortstack[c]{\textbf{$b$}\\\textbf{(SE)}} & \shortstack[c]{\textbf{$t(df)$}\\\textbf{$p$}} & \shortstack[c]{\textbf{$b$}\\\textbf{(SE)}} & \shortstack[c]{\textbf{$t(df)$}\\\textbf{$p$}} & \shortstack[c]{\textbf{$b$}\\\textbf{(SE)}} & \shortstack[c]{\textbf{$t(df)$}\\\textbf{$p$}} & \shortstack[c]{\textbf{$b$}\\\textbf{(SE)}} & \shortstack[c]{\textbf{$t(df)$}\\\textbf{$p$}} & \shortstack[c]{\textbf{$b$}\\\textbf{(SE)}} & \shortstack[c]{\textbf{$t(df)$}\\\textbf{$p$}} \\
\addlinespace[0.25em]
(Intercept) & \shortstack[c]{0.731***\\(0.009)} & \shortstack[c]{t(413)=81.12\\p<.001} & \shortstack[c]{0.685***\\(0.019)} & \shortstack[c]{t(393)=36.49\\p<.001} & \shortstack[c]{0.752***\\(0.019)} & \shortstack[c]{t(393)=39.85\\p<.001} & \shortstack[c]{0.376***\\(0.048)} & \shortstack[c]{t(408)=7.92\\p<.001} & \shortstack[c]{0.365***\\(0.049)} & \shortstack[c]{t(388)=7.51\\p<.001} & \shortstack[c]{0.409***\\(0.052)} & \shortstack[c]{t(388)=7.90\\p<.001} \\
Experimental Manipulation (LC) &  &  & \shortstack[c]{0.068*\\(0.026)} & \shortstack[c]{t(16)=2.58\\p=0.020} &  &  &  &  & \shortstack[c]{0.044\\(0.026)} & \shortstack[c]{t(16)=1.67\\p=0.115} &  &  \\
Experimental Manipulation (PE) &  &  & \shortstack[c]{0.036\\(0.026)} & \shortstack[c]{t(16)=1.40\\p=0.180} & \shortstack[c]{-0.031\\(0.026)} & \shortstack[c]{t(16)=-1.20\\p=0.249} &  &  & \shortstack[c]{0.024\\(0.026)} & \shortstack[c]{t(16)=0.91\\p=0.377} & \shortstack[c]{-0.020\\(0.026)} & \shortstack[c]{t(16)=-0.77\\p=0.451} \\
Experimental Manipulation (NE) &  &  & \shortstack[c]{0.054\\(0.026)} & \shortstack[c]{t(16)=2.06\\p=0.056} & \shortstack[c]{-0.013\\(0.026)} & \shortstack[c]{t(16)=-0.51\\p=0.620} &  &  & \shortstack[c]{0.025\\(0.026)} & \shortstack[c]{t(16)=0.95\\p=0.357} & \shortstack[c]{-0.019\\(0.026)} & \shortstack[c]{t(16)=-0.71\\p=0.488} \\
Experimental Manipulation (FD) &  &  & \shortstack[c]{0.073*\\(0.026)} & \shortstack[c]{t(16)=2.80\\p=0.013} & \shortstack[c]{0.006\\(0.026)} & \shortstack[c]{t(16)=0.22\\p=0.826} &  &  & \shortstack[c]{0.041\\(0.027)} & \shortstack[c]{t(16)=1.55\\p=0.140} & \shortstack[c]{-0.003\\(0.026)} & \shortstack[c]{t(16)=-0.10\\p=0.923} \\
Experimental Manipulation (LI) &  &  &  &  & \shortstack[c]{-0.068*\\(0.026)} & \shortstack[c]{t(16)=-2.58\\p=0.020} &  &  &  &  & \shortstack[c]{-0.044\\(0.026)} & \shortstack[c]{t(16)=-1.67\\p=0.115} \\
Effort &  &  &  &  &  &  & \shortstack[c]{0.001\\(0.001)} & \shortstack[c]{t(408)=1.26\\p=0.207} & \shortstack[c]{0.001\\(0.001)} & \shortstack[c]{t(388)=1.07\\p=0.285} & \shortstack[c]{0.001\\(0.001)} & \shortstack[c]{t(388)=1.07\\p=0.285} \\
Familiarity &  &  &  &  &  &  & \shortstack[c]{0.008\\(0.009)} & \shortstack[c]{t(408)=0.86\\p=0.391} & \shortstack[c]{0.007\\(0.009)} & \shortstack[c]{t(388)=0.80\\p=0.422} & \shortstack[c]{0.007\\(0.009)} & \shortstack[c]{t(388)=0.80\\p=0.422} \\
AI-Knowledge &  &  &  &  &  &  & \shortstack[c]{0.001\\(0.006)} & \shortstack[c]{t(408)=0.21\\p=0.833} & \shortstack[c]{0.001\\(0.006)} & \shortstack[c]{t(388)=0.13\\p=0.899} & \shortstack[c]{0.001\\(0.006)} & \shortstack[c]{t(388)=0.13\\p=0.899} \\
Skin Cancer Knowledge &  &  &  &  &  &  & \shortstack[c]{0.008\\(0.005)} & \shortstack[c]{t(408)=1.54\\p=0.125} & \shortstack[c]{0.007\\(0.005)} & \shortstack[c]{t(388)=1.38\\p=0.168} & \shortstack[c]{0.007\\(0.005)} & \shortstack[c]{t(388)=1.38\\p=0.168} \\
Propensity To Trust &  &  &  &  &  &  & \shortstack[c]{0.092***\\(0.011)} & \shortstack[c]{t(408)=8.19\\p<.001} & \shortstack[c]{0.090***\\(0.011)} & \shortstack[c]{t(388)=7.94\\p<.001} & \shortstack[c]{0.090***\\(0.011)} & \shortstack[c]{t(388)=7.94\\p<.001} \\
\addlinespace[0.6em]
\midrule
\textbf{Model fit} & \multicolumn{2}{c}{\shortstack[c]{\textbf{AIC}\\\textbf{Log Likelihood}}} & \multicolumn{4}{c}{\shortstack[c]{\textbf{AIC}\\\textbf{Log Likelihood}}} & \multicolumn{2}{c}{\shortstack[c]{\textbf{AIC}\\\textbf{Log Likelihood}}} & \multicolumn{4}{c}{\shortstack[c]{\textbf{AIC}\\\textbf{Log Likelihood}}} \\
\addlinespace[0.25em]
 & \multicolumn{2}{c}{\shortstack[c]{-287.017\\146.508}} & \multicolumn{4}{c}{\shortstack[c]{-287.220\\151.610}} & \multicolumn{2}{c}{\shortstack[c]{-349.976\\182.988}} & \multicolumn{4}{c}{\shortstack[c]{-344.314\\185.157}} \\
\textit{Likelihood Ratio Test vs. prev. Model} & \multicolumn{2}{c}{} & \multicolumn{4}{c}{} & \multicolumn{2}{c}{} & \multicolumn{4}{c}{} \\
\addlinespace[0.6em]
\midrule
\textbf{Variance explained (R\textsuperscript{2} in \%)} & \multicolumn{2}{c}{\shortstack[c]{\textbf{Marginal $R^2$}\\\textbf{Conditional $R^2$}}} & \multicolumn{4}{c}{\shortstack[c]{\textbf{Marginal $R^2$}\\\textbf{Conditional $R^2$}}} & \multicolumn{2}{c}{\shortstack[c]{\textbf{Marginal $R^2$}\\\textbf{Conditional $R^2$}}} & \multicolumn{4}{c}{\shortstack[c]{\textbf{Marginal $R^2$}\\\textbf{Conditional $R^2$}}} \\
\addlinespace[0.25em]
 & \multicolumn{2}{c}{\shortstack[c]{0.0\\0.2}} & \multicolumn{4}{c}{\shortstack[c]{2.4\\2.6}} & \multicolumn{2}{c}{\shortstack[c]{16.1\\16.6}} & \multicolumn{4}{c}{\shortstack[c]{17.0\\18.0}} \\
\bottomrule
\end{tabular}

    }
\end{sidewaystable}
\clearpage

\begin{sidewaystable}[p]
    \centering
    \caption{Linear Mixed Effect Models predicting the combined on-task trust/trustworthiness measure}
    \label{tab:ontasktrust_overall_combined}

    \adjustbox{max width=\textheight, max totalheight=\linewidth, keepaspectratio}{
        \begin{tabular}{lcccccccccccc}
\toprule
\textbf{Model component} & \multicolumn{2}{c}{\textbf{onTask-M1: Only Packages}} & \multicolumn{2}{c}{\textbf{onTask-M2a: NE vs Others}} & \multicolumn{2}{c}{\textbf{onTask-M2b: LC vs Others}} & \multicolumn{2}{c}{\textbf{onTask-M3: Control + Packages}} & \multicolumn{2}{c}{\textbf{onTask-M4a: NE vs Others}} & \multicolumn{2}{c}{\textbf{onTask-M4b: LC vs Others}} \\
\cmidrule(lr){2-3}\cmidrule(lr){4-5}\cmidrule(lr){6-7}\cmidrule(lr){8-9}\cmidrule(lr){10-11}\cmidrule(lr){12-13}
\midrule
\textbf{Random Effects} & \multicolumn{2}{c}{\textbf{SD}} & \multicolumn{4}{c}{\textbf{SD}} & \multicolumn{2}{c}{\textbf{SD}} & \multicolumn{4}{c}{\textbf{SD}} \\
\addlinespace[0.25em]
Stimulus Package (intercept) & \multicolumn{2}{c}{0.032} & \multicolumn{4}{c}{0.031} & \multicolumn{2}{c}{0.033} & \multicolumn{4}{c}{0.032} \\
Experimental Manipulation (intercept) & \multicolumn{2}{c}{} & \multicolumn{4}{c}{0.021} & \multicolumn{2}{c}{} & \multicolumn{4}{c}{0.022} \\
Residual & \multicolumn{2}{c}{0.122} & \multicolumn{4}{c}{0.119} & \multicolumn{2}{c}{0.119} & \multicolumn{4}{c}{0.116} \\
\addlinespace[0.6em]
\midrule
\textbf{Fixed Effects} & \shortstack[c]{\textbf{$b$}\\\textbf{(SE)}} & \shortstack[c]{\textbf{$t(df)$}\\\textbf{$p$}} & \shortstack[c]{\textbf{$b$}\\\textbf{(SE)}} & \shortstack[c]{\textbf{$t(df)$}\\\textbf{$p$}} & \shortstack[c]{\textbf{$b$}\\\textbf{(SE)}} & \shortstack[c]{\textbf{$t(df)$}\\\textbf{$p$}} & \shortstack[c]{\textbf{$b$}\\\textbf{(SE)}} & \shortstack[c]{\textbf{$t(df)$}\\\textbf{$p$}} & \shortstack[c]{\textbf{$b$}\\\textbf{(SE)}} & \shortstack[c]{\textbf{$t(df)$}\\\textbf{$p$}} & \shortstack[c]{\textbf{$b$}\\\textbf{(SE)}} & \shortstack[c]{\textbf{$t(df)$}\\\textbf{$p$}} \\
\addlinespace[0.25em]
(Intercept) & \shortstack[c]{0.703***\\(0.015)} & \shortstack[c]{t(413)=45.52\\p<.001} & \shortstack[c]{0.668***\\(0.021)} & \shortstack[c]{t(393)=31.05\\p<.001} & \shortstack[c]{0.731***\\(0.022)} & \shortstack[c]{t(393)=33.83\\p<.001} & \shortstack[c]{0.600***\\(0.039)} & \shortstack[c]{t(408)=15.38\\p<.001} & \shortstack[c]{0.585***\\(0.040)} & \shortstack[c]{t(388)=14.50\\p<.001} & \shortstack[c]{0.642***\\(0.043)} & \shortstack[c]{t(388)=15.08\\p<.001} \\
Experimental Manipulation (LC) &  &  & \shortstack[c]{0.063*\\(0.023)} & \shortstack[c]{t(16)=2.73\\p=0.015} &  &  &  &  & \shortstack[c]{0.057*\\(0.023)} & \shortstack[c]{t(16)=2.43\\p=0.027} &  &  \\
Experimental Manipulation (PE) &  &  & \shortstack[c]{0.030\\(0.023)} & \shortstack[c]{t(16)=1.31\\p=0.210} & \shortstack[c]{-0.033\\(0.023)} & \shortstack[c]{t(16)=-1.43\\p=0.172} &  &  & \shortstack[c]{0.029\\(0.023)} & \shortstack[c]{t(16)=1.24\\p=0.231} & \shortstack[c]{-0.028\\(0.023)} & \shortstack[c]{t(16)=-1.20\\p=0.248} \\
Experimental Manipulation (NE) &  &  & \shortstack[c]{0.045\\(0.023)} & \shortstack[c]{t(16)=1.93\\p=0.071} & \shortstack[c]{-0.018\\(0.023)} & \shortstack[c]{t(16)=-0.78\\p=0.447} &  &  & \shortstack[c]{0.034\\(0.024)} & \shortstack[c]{t(16)=1.45\\p=0.167} & \shortstack[c]{-0.023\\(0.023)} & \shortstack[c]{t(16)=-0.97\\p=0.345} \\
Experimental Manipulation (FD) &  &  & \shortstack[c]{0.044\\(0.023)} & \shortstack[c]{t(16)=1.90\\p=0.076} & \shortstack[c]{-0.019\\(0.023)} & \shortstack[c]{t(16)=-0.83\\p=0.421} &  &  & \shortstack[c]{0.035\\(0.024)} & \shortstack[c]{t(16)=1.49\\p=0.156} & \shortstack[c]{-0.022\\(0.023)} & \shortstack[c]{t(16)=-0.93\\p=0.365} \\
Experimental Manipulation (LI) &  &  &  &  & \shortstack[c]{-0.063*\\(0.023)} & \shortstack[c]{t(16)=-2.73\\p=0.015} &  &  &  &  & \shortstack[c]{-0.057*\\(0.023)} & \shortstack[c]{t(16)=-2.43\\p=0.027} \\
Effort &  &  &  &  &  &  & \shortstack[c]{0.001\\(0.001)} & \shortstack[c]{t(408)=1.21\\p=0.226} & \shortstack[c]{0.001\\(0.001)} & \shortstack[c]{t(388)=1.04\\p=0.300} & \shortstack[c]{0.001\\(0.001)} & \shortstack[c]{t(388)=1.04\\p=0.300} \\
Familiarity &  &  &  &  &  &  & \shortstack[c]{-0.014*\\(0.007)} & \shortstack[c]{t(408)=-2.03\\p=0.043} & \shortstack[c]{-0.014*\\(0.007)} & \shortstack[c]{t(388)=-2.17\\p=0.031} & \shortstack[c]{-0.014*\\(0.007)} & \shortstack[c]{t(388)=-2.17\\p=0.031} \\
AI-Knowledge &  &  &  &  &  &  & \shortstack[c]{-0.003\\(0.005)} & \shortstack[c]{t(408)=-0.68\\p=0.497} & \shortstack[c]{-0.004\\(0.005)} & \shortstack[c]{t(388)=-0.79\\p=0.427} & \shortstack[c]{-0.004\\(0.005)} & \shortstack[c]{t(388)=-0.79\\p=0.427} \\
Skin Cancer Knowledge &  &  &  &  &  &  & \shortstack[c]{0.005\\(0.004)} & \shortstack[c]{t(408)=1.39\\p=0.165} & \shortstack[c]{0.005\\(0.004)} & \shortstack[c]{t(388)=1.29\\p=0.196} & \shortstack[c]{0.005\\(0.004)} & \shortstack[c]{t(388)=1.29\\p=0.196} \\
Propensity To Trust &  &  &  &  &  &  & \shortstack[c]{0.032***\\(0.009)} & \shortstack[c]{t(408)=3.69\\p<.001} & \shortstack[c]{0.030***\\(0.009)} & \shortstack[c]{t(388)=3.41\\p<.001} & \shortstack[c]{0.030***\\(0.009)} & \shortstack[c]{t(388)=3.41\\p<.001} \\
\addlinespace[0.6em]
\midrule
\textbf{Model fit} & \multicolumn{2}{c}{\shortstack[c]{\textbf{AIC}\\\textbf{Log Likelihood}}} & \multicolumn{4}{c}{\shortstack[c]{\textbf{AIC}\\\textbf{Log Likelihood}}} & \multicolumn{2}{c}{\shortstack[c]{\textbf{AIC}\\\textbf{Log Likelihood}}} & \multicolumn{4}{c}{\shortstack[c]{\textbf{AIC}\\\textbf{Log Likelihood}}} \\
\addlinespace[0.25em]
 & \multicolumn{2}{c}{\shortstack[c]{-556.689\\281.345}} & \multicolumn{4}{c}{\shortstack[c]{-561.003\\288.502}} & \multicolumn{2}{c}{\shortstack[c]{-567.040\\291.520}} & \multicolumn{4}{c}{\shortstack[c]{-569.232\\297.616}} \\
\textit{Likelihood Ratio Test vs. prev. Model} & \multicolumn{2}{c}{} & \multicolumn{4}{c}{*} & \multicolumn{2}{c}{} & \multicolumn{4}{c}{*} \\
\addlinespace[0.6em]
\midrule
\textbf{Variance explained (R\textsuperscript{2} in \%)} & \multicolumn{2}{c}{\shortstack[c]{\textbf{Marginal $R^2$}\\\textbf{Conditional $R^2$}}} & \multicolumn{4}{c}{\shortstack[c]{\textbf{Marginal $R^2$}\\\textbf{Conditional $R^2$}}} & \multicolumn{2}{c}{\shortstack[c]{\textbf{Marginal $R^2$}\\\textbf{Conditional $R^2$}}} & \multicolumn{4}{c}{\shortstack[c]{\textbf{Marginal $R^2$}\\\textbf{Conditional $R^2$}}} \\
\addlinespace[0.25em]
 & \multicolumn{2}{c}{\shortstack[c]{0.0\\6.4}} & \multicolumn{4}{c}{\shortstack[c]{2.8\\11.8}} & \multicolumn{2}{c}{\shortstack[c]{4.5\\11.2}} & \multicolumn{4}{c}{\shortstack[c]{6.6\\16.1}} \\
\bottomrule
\end{tabular}

    }
\end{sidewaystable}
\clearpage

\begin{sidewaystable}[p]
    \centering
    \caption{Linear Mixed Effect Models predicting the accuracy estimation measure}
    \label{tab:estimatedAccuracy_overall_combined}

    \adjustbox{max width=\textheight, max totalheight=\linewidth, keepaspectratio}{
        \begin{tabular}{lcccccccccccc}
\toprule
\textbf{Model component} & \multicolumn{2}{c}{\textbf{accEstimation-M1: Only Packages}} & \multicolumn{2}{c}{\textbf{accEstimation-M2a: NE vs Others}} & \multicolumn{2}{c}{\textbf{accEstimation-M2b: LC vs Others}} & \multicolumn{2}{c}{\textbf{accEstimation-M3: Control + Packages}} & \multicolumn{2}{c}{\textbf{accEstimation-M4a: NE vs Others}} & \multicolumn{2}{c}{\textbf{accEstimation-M4b: LC vs Others}} \\
\cmidrule(lr){2-3}\cmidrule(lr){4-5}\cmidrule(lr){6-7}\cmidrule(lr){8-9}\cmidrule(lr){10-11}\cmidrule(lr){12-13}
\midrule
\textbf{Random Effects} & \multicolumn{2}{c}{\textbf{SD}} & \multicolumn{4}{c}{\textbf{SD}} & \multicolumn{2}{c}{\textbf{SD}} & \multicolumn{4}{c}{\textbf{SD}} \\
\addlinespace[0.25em]
Stimulus Package (intercept) & \multicolumn{2}{c}{0.030} & \multicolumn{4}{c}{0.031} & \multicolumn{2}{c}{0.031} & \multicolumn{4}{c}{0.031} \\
Experimental Manipulation (intercept) & \multicolumn{2}{c}{} & \multicolumn{4}{c}{0.015} & \multicolumn{2}{c}{} & \multicolumn{4}{c}{0.016} \\
Residual & \multicolumn{2}{c}{0.112} & \multicolumn{4}{c}{0.109} & \multicolumn{2}{c}{0.110} & \multicolumn{4}{c}{0.107} \\
\addlinespace[0.6em]
\midrule
\textbf{Fixed Effects} & \shortstack[c]{\textbf{$b$}\\\textbf{(SE)}} & \shortstack[c]{\textbf{$t(df)$}\\\textbf{$p$}} & \shortstack[c]{\textbf{$b$}\\\textbf{(SE)}} & \shortstack[c]{\textbf{$t(df)$}\\\textbf{$p$}} & \shortstack[c]{\textbf{$b$}\\\textbf{(SE)}} & \shortstack[c]{\textbf{$t(df)$}\\\textbf{$p$}} & \shortstack[c]{\textbf{$b$}\\\textbf{(SE)}} & \shortstack[c]{\textbf{$t(df)$}\\\textbf{$p$}} & \shortstack[c]{\textbf{$b$}\\\textbf{(SE)}} & \shortstack[c]{\textbf{$t(df)$}\\\textbf{$p$}} & \shortstack[c]{\textbf{$b$}\\\textbf{(SE)}} & \shortstack[c]{\textbf{$t(df)$}\\\textbf{$p$}} \\
\addlinespace[0.25em]
(Intercept) & \shortstack[c]{0.720***\\(0.015)} & \shortstack[c]{t(413)=49.49\\p<.001} & \shortstack[c]{0.693***\\(0.020)} & \shortstack[c]{t(393)=35.50\\p<.001} & \shortstack[c]{0.752***\\(0.020)} & \shortstack[c]{t(393)=38.36\\p<.001} & \shortstack[c]{0.663***\\(0.036)} & \shortstack[c]{t(408)=18.32\\p<.001} & \shortstack[c]{0.653***\\(0.037)} & \shortstack[c]{t(388)=17.57\\p<.001} & \shortstack[c]{0.709***\\(0.039)} & \shortstack[c]{t(388)=18.08\\p<.001} \\
Experimental Manipulation (LC) &  &  & \shortstack[c]{0.059**\\(0.019)} & \shortstack[c]{t(16)=3.02\\p=0.008} &  &  &  &  & \shortstack[c]{0.056*\\(0.020)} & \shortstack[c]{t(16)=2.80\\p=0.013} &  &  \\
Experimental Manipulation (PE) &  &  & \shortstack[c]{0.012\\(0.019)} & \shortstack[c]{t(16)=0.62\\p=0.547} & \shortstack[c]{-0.047*\\(0.019)} & \shortstack[c]{t(16)=-2.40\\p=0.029} &  &  & \shortstack[c]{0.012\\(0.020)} & \shortstack[c]{t(16)=0.61\\p=0.551} & \shortstack[c]{-0.044*\\(0.020)} & \shortstack[c]{t(16)=-2.20\\p=0.043} \\
Experimental Manipulation (NE) &  &  & \shortstack[c]{0.035\\(0.020)} & \shortstack[c]{t(16)=1.79\\p=0.092} & \shortstack[c]{-0.024\\(0.020)} & \shortstack[c]{t(16)=-1.21\\p=0.245} &  &  & \shortstack[c]{0.029\\(0.020)} & \shortstack[c]{t(16)=1.45\\p=0.165} & \shortstack[c]{-0.027\\(0.020)} & \shortstack[c]{t(16)=-1.33\\p=0.203} \\
Experimental Manipulation (FD) &  &  & \shortstack[c]{0.032\\(0.019)} & \shortstack[c]{t(16)=1.64\\p=0.119} & \shortstack[c]{-0.027\\(0.020)} & \shortstack[c]{t(16)=-1.36\\p=0.192} &  &  & \shortstack[c]{0.027\\(0.020)} & \shortstack[c]{t(16)=1.32\\p=0.206} & \shortstack[c]{-0.029\\(0.020)} & \shortstack[c]{t(16)=-1.46\\p=0.163} \\
Experimental Manipulation (LI) &  &  &  &  & \shortstack[c]{-0.059**\\(0.019)} & \shortstack[c]{t(16)=-3.02\\p=0.008} &  &  &  &  & \shortstack[c]{-0.056*\\(0.020)} & \shortstack[c]{t(16)=-2.80\\p=0.013} \\
Effort &  &  &  &  &  &  & \shortstack[c]{0.001\\(0.000)} & \shortstack[c]{t(408)=1.29\\p=0.196} & \shortstack[c]{0.000\\(0.000)} & \shortstack[c]{t(388)=1.00\\p=0.320} & \shortstack[c]{0.000\\(0.000)} & \shortstack[c]{t(388)=1.00\\p=0.320} \\
Familiarity &  &  &  &  &  &  & \shortstack[c]{-0.013*\\(0.006)} & \shortstack[c]{t(408)=-2.08\\p=0.038} & \shortstack[c]{-0.014*\\(0.006)} & \shortstack[c]{t(388)=-2.19\\p=0.029} & \shortstack[c]{-0.014*\\(0.006)} & \shortstack[c]{t(388)=-2.19\\p=0.029} \\
AI-Knowledge &  &  &  &  &  &  & \shortstack[c]{-0.003\\(0.004)} & \shortstack[c]{t(408)=-0.66\\p=0.508} & \shortstack[c]{-0.003\\(0.004)} & \shortstack[c]{t(388)=-0.73\\p=0.467} & \shortstack[c]{-0.003\\(0.004)} & \shortstack[c]{t(388)=-0.73\\p=0.467} \\
Skin Cancer Knowledge &  &  &  &  &  &  & \shortstack[c]{0.006\\(0.004)} & \shortstack[c]{t(408)=1.60\\p=0.110} & \shortstack[c]{0.006\\(0.004)} & \shortstack[c]{t(388)=1.58\\p=0.115} & \shortstack[c]{0.006\\(0.004)} & \shortstack[c]{t(388)=1.58\\p=0.115} \\
Propensity To Trust &  &  &  &  &  &  & \shortstack[c]{0.016*\\(0.008)} & \shortstack[c]{t(408)=2.01\\p=0.045} & \shortstack[c]{0.014\\(0.008)} & \shortstack[c]{t(388)=1.69\\p=0.092} & \shortstack[c]{0.014\\(0.008)} & \shortstack[c]{t(388)=1.69\\p=0.092} \\
\addlinespace[0.6em]
\midrule
\textbf{Model fit} & \multicolumn{2}{c}{\shortstack[c]{\textbf{AIC}\\\textbf{Log Likelihood}}} & \multicolumn{4}{c}{\shortstack[c]{\textbf{AIC}\\\textbf{Log Likelihood}}} & \multicolumn{2}{c}{\shortstack[c]{\textbf{AIC}\\\textbf{Log Likelihood}}} & \multicolumn{4}{c}{\shortstack[c]{\textbf{AIC}\\\textbf{Log Likelihood}}} \\
\addlinespace[0.25em]
 & \multicolumn{2}{c}{\shortstack[c]{-631.147\\318.573}} & \multicolumn{4}{c}{\shortstack[c]{-635.601\\325.800}} & \multicolumn{2}{c}{\shortstack[c]{-632.748\\324.374}} & \multicolumn{4}{c}{\shortstack[c]{-635.681\\330.841}} \\
\textit{Likelihood Ratio Test vs. prev. Model} & \multicolumn{2}{c}{} & \multicolumn{4}{c}{*} & \multicolumn{2}{c}{} & \multicolumn{4}{c}{*} \\
\addlinespace[0.6em]
\midrule
\textbf{Variance explained (R\textsuperscript{2} in \%)} & \multicolumn{2}{c}{\shortstack[c]{\textbf{Marginal $R^2$}\\\textbf{Conditional $R^2$}}} & \multicolumn{4}{c}{\shortstack[c]{\textbf{Marginal $R^2$}\\\textbf{Conditional $R^2$}}} & \multicolumn{2}{c}{\shortstack[c]{\textbf{Marginal $R^2$}\\\textbf{Conditional $R^2$}}} & \multicolumn{4}{c}{\shortstack[c]{\textbf{Marginal $R^2$}\\\textbf{Conditional $R^2$}}} \\
\addlinespace[0.25em]
 & \multicolumn{2}{c}{\shortstack[c]{0.0\\6.8}} & \multicolumn{4}{c}{\shortstack[c]{3.1\\11.8}} & \multicolumn{2}{c}{\shortstack[c]{2.6\\9.7}} & \multicolumn{4}{c}{\shortstack[c]{5.2\\14.5}} \\
\bottomrule
\end{tabular}

    }
\end{sidewaystable}
\clearpage

\begin{sidewaystable}[p]
    \centering
    \caption{Linear Mixed Effect Models predicting the combined on-task trust/trustworthiness measure over time}
    \label{tab:ontasktrust_overTime}

    \adjustbox{max width=\textheight, max totalheight=\linewidth, keepaspectratio}{
        \begin{tabular}{lcccccccc}
\toprule
\textbf{Model component} & \multicolumn{2}{c}{\textbf{onTask-Longitudinal-M1a: LI vs Others}} & \multicolumn{2}{c}{\textbf{onTask-Longitudinal-M1b: LC vs Others}} & \multicolumn{2}{c}{\textbf{onTask-Longitudinal-M2a: Mediation}} & \multicolumn{2}{c}{\textbf{onTask-Longitudinal-M2b: Mediation}} \\
\cmidrule(lr){2-3}\cmidrule(lr){4-5}\cmidrule(lr){6-7}\cmidrule(lr){8-9}
\midrule
\textbf{Random Effects} & \multicolumn{4}{c}{\textbf{SD}} & \multicolumn{4}{c}{\textbf{SD}} \\
\addlinespace[0.25em]
Participant ID (intercept) & \multicolumn{4}{c}{0.094} & \multicolumn{4}{c}{0.094} \\
ParticipantID (slope of Stimulus Order) & \multicolumn{4}{c}{0.001} & \multicolumn{4}{c}{0.001} \\
Stimulus Order (intercept) & \multicolumn{4}{c}{<.001} & \multicolumn{4}{c}{<.001} \\
Experimental Manipulation (intercept) & \multicolumn{4}{c}{0.019} & \multicolumn{4}{c}{0.252} \\
Stimulus Package (intercept) & \multicolumn{4}{c}{0.037} & \multicolumn{4}{c}{0.037} \\
Residual & \multicolumn{4}{c}{0.241} & \multicolumn{4}{c}{0.241} \\
Residual & \multicolumn{4}{c}{} & \multicolumn{4}{c}{} \\
\addlinespace[0.6em]
\midrule
\textbf{Fixed Effects} & \shortstack[c]{\textbf{$b$}\\\textbf{(SE)}} & \shortstack[c]{\textbf{$t(df)$}\\\textbf{$p$}} & \shortstack[c]{\textbf{$b$}\\\textbf{(SE)}} & \shortstack[c]{\textbf{$t(df)$}\\\textbf{$p$}} & \shortstack[c]{\textbf{$b$}\\\textbf{(SE)}} & \shortstack[c]{\textbf{$t(df)$}\\\textbf{$p$}} & \shortstack[c]{\textbf{$b$}\\\textbf{(SE)}} & \shortstack[c]{\textbf{$t(df)$}\\\textbf{$p$}} \\
\addlinespace[0.25em]
(Intercept) & \shortstack[c]{0.718***\\(0.020)} & \shortstack[c]{t(6)=35.55\\p<.001} & \shortstack[c]{0.718***\\(0.020)} & \shortstack[c]{t(6)=35.55\\p<.001} & \shortstack[c]{0.749\\(0.253)} & \shortstack[c]{t(0)=2.96\\p=1.000} & \shortstack[c]{0.749\\(0.253)} & \shortstack[c]{t(0)=2.96\\p=1.000} \\
Stimulus Order & \shortstack[c]{-0.002**\\(0.001)} & \shortstack[c]{t(672)=-2.71\\p=0.007} & \shortstack[c]{-0.002**\\(0.001)} & \shortstack[c]{t(672)=-2.71\\p=0.007} & \shortstack[c]{-0.002\\(0.002)} & \shortstack[c]{t(4696)=-1.50\\p=0.134} & \shortstack[c]{-0.002\\(0.002)} & \shortstack[c]{t(4696)=-1.50\\p=0.134} \\
Experimental Manipulation (LI) &  &  &  &  & \shortstack[c]{-0.065\\(0.357)} & \shortstack[c]{t(0)=-0.18\\p=1.000} & \shortstack[c]{-0.065\\(0.357)} & \shortstack[c]{t(0)=-0.18\\p=1.000} \\
Experimental Manipulation (PE) &  &  &  &  & \shortstack[c]{-0.032\\(0.357)} & \shortstack[c]{t(0)=-0.09\\p=1.000} & \shortstack[c]{-0.032\\(0.357)} & \shortstack[c]{t(0)=-0.09\\p=1.000} \\
Experimental Manipulation (NE) &  &  &  &  & \shortstack[c]{-0.015\\(0.357)} & \shortstack[c]{t(0)=-0.04\\p=1.000} & \shortstack[c]{-0.015\\(0.357)} & \shortstack[c]{t(0)=-0.04\\p=1.000} \\
Experimental Manipulation (FD) &  &  &  &  & \shortstack[c]{-0.039\\(0.357)} & \shortstack[c]{t(0)=-0.11\\p=1.000} & \shortstack[c]{-0.039\\(0.357)} & \shortstack[c]{t(0)=-0.11\\p=1.000} \\
Stimulus Order:Experimental Manipulation (LI) &  &  &  &  & \shortstack[c]{0.000\\(0.002)} & \shortstack[c]{t(4740)=0.08\\p=0.933} & \shortstack[c]{0.000\\(0.002)} & \shortstack[c]{t(4740)=0.08\\p=0.933} \\
Stimulus Order:Experimental Manipulation (PE) &  &  &  &  & \shortstack[c]{-0.000\\(0.002)} & \shortstack[c]{t(4740)=-0.02\\p=0.983} & \shortstack[c]{-0.000\\(0.002)} & \shortstack[c]{t(4740)=-0.02\\p=0.983} \\
Stimulus Order:Experimental Manipulation (NE) &  &  &  &  & \shortstack[c]{-0.000\\(0.002)} & \shortstack[c]{t(4740)=-0.17\\p=0.865} & \shortstack[c]{-0.000\\(0.002)} & \shortstack[c]{t(4740)=-0.17\\p=0.865} \\
Stimulus Order:Experimental Manipulation (FD) &  &  &  &  & \shortstack[c]{0.003\\(0.002)} & \shortstack[c]{t(4740)=1.14\\p=0.253} & \shortstack[c]{0.003\\(0.002)} & \shortstack[c]{t(4740)=1.14\\p=0.253} \\
\addlinespace[0.6em]
\midrule
\textbf{Model fit} & \multicolumn{4}{c}{\shortstack[c]{\textbf{AIC}\\\textbf{Log Likelihood}}} & \multicolumn{4}{c}{\shortstack[c]{\textbf{AIC}\\\textbf{Log Likelihood}}} \\
\addlinespace[0.25em]
 & \multicolumn{4}{c}{\shortstack[c]{565.123\\-273.562}} & \multicolumn{4}{c}{\shortstack[c]{638.564\\-302.282}} \\
\textit{Likelihood Ratio Test vs. prev. Model} & \multicolumn{4}{c}{} & \multicolumn{4}{c}{} \\
\addlinespace[0.6em]
\midrule
\textbf{Variance explained (R\textsuperscript{2} in \%)} & \multicolumn{4}{c}{\shortstack[c]{\textbf{Marginal $R^2$}\\\textbf{Conditional $R^2$}}} & \multicolumn{4}{c}{\shortstack[c]{\textbf{Marginal $R^2$}\\\textbf{Conditional $R^2$}}} \\
\addlinespace[0.25em]
 & \multicolumn{4}{c}{\shortstack[c]{<0.1\\17.7}} & \multicolumn{4}{c}{\shortstack[c]{0.4\\56.6}} \\
\bottomrule
\end{tabular}

    }
\end{sidewaystable}
\clearpage

\end{document}